\documentclass[epj]{svjour}

\usepackage{amsmath}
\usepackage{amssymb}
\usepackage[sort&compress]{natbib}
\usepackage{enumerate}
\usepackage{graphicx}
\usepackage{dcolumn}
\usepackage{bm}
\usepackage{wasysym}
\usepackage{ae,aecompl}

\usepackage[english]{babel}
\usepackage[latin2]{inputenc}
\usepackage[T1]{fontenc}

\def\ccc#1;#2{\left\langle #1 \left\vert #2 \right.\right\rangle}
\def\ev #1{\left\langle #1 \right\rangle}

\begin{document}

\title{Size matters: some stylized facts of the stock market revisited}
\author{Zolt\'an Eisler\inst{1}\thanks{\email{eisler@maxwell.phy.bme.hu}} \and J\'anos Kert\'esz\inst{1,2}}
\institute{
  \inst{1} Department of Theoretical Physics, Budapest University of
  Technology and Economics - Budapest, Hungary\\
  \inst{2} Laboratory of Computational Engineering,
  Helsinki University of Technology - Espoo, Finland}

\date{\today}

\abstract{We reanalyze high resolution data from the New York Stock
Exchange and find a monotonic (but not power law) variation of the mean value per trade,
the mean number of trades per minute and the mean trading activity
with company capitalization. We show that the second moment of the
traded value distribution is finite. Consequently, the Hurst
exponents for the corresponding time series can be calculated. These
are, however, non-universal: The persistence grows with larger
capitalization and this results in a logarithmically increasing Hurst exponent. A similar
trend is displayed by intertrade time intervals.
Finally, we demonstrate that the distribution of the intertrade times is
better described by a multiscaling ansatz than by simple gap scaling.}


\maketitle

Understanding the financial market as a self-adaptive, strongly
interacting system is a real interdisciplinary challenge, where
physicists strongly hope to make essential contributions
\cite{evolving,evolving2,kertesz.econophysics}.  The enthusiasm is
understandable as the breakthrough of the early $70$'s in statistical
physics taught us how to handle strongly interacting systems with a
large number of degrees of freedom. The unbroken development of this
and related disciplines brought up several concepts and models like
(fractal and multifractal) scaling, frustrated disordered systems, or
far from equilibrium phenomena and we have obtained very efficient
tools to treat them. Many of us are convinced, that these and similar
ideas and techniques will be helpful to understand the mechanisms of
the economy. In fact, there have been quite successful attempts along
this line \cite{bouchaud.book, stanley.book, mandelbrot.econophysics}.
An ubiquitous aspect of strongly interacting systems is the lack of
finite scales. The best understood examples are second order
equilibrium phase transitions where renormalization group theory
provides a general explanation of scaling and universality
\cite{reichl}.  It seems that some features of the stock market can
indeed be captured by these concepts: For example, the so called
inverse cube law of the distribution of logarithmic returns shows a
quite convincing data collapse for different companies with a good fit
to an algebraically decaying tail \cite{gopi.inversecube,
lux.paretian}.

Studies in econophysics concentrate on the possible analogies,
although there are important differences between physical and
financial systems. This is, of course, a trivial statement -- it is
enough to refer to the above-mentioned self-adaptivity, to the
possibility of influencing the system by its characterization or to
the intrinsic non-stationarity of economic processes. Here we would
like to emphasize the discrepancy in the levels of description. In the
case of a physical system undergoing a second order phase transition,
it is natural to assume scaling on profound theoretical grounds and
the (experimental or theoretical) determination of, e.g., the critical
exponents is a fully justified undertaking. There is no similar
theoretical basis for the financial market whatsoever, therefore in
this case the assumption of power laws should be considered only as
one possible way of fitting fat tailed distributions. Also, the
reference to universality should not be plausible as the robustness of
\emph{qualitative} features -- like the fat tail of the distributions
-- is a much weaker property.  Therefore, e.g., averaging
distributions over companies with very different capitalization is
questionable. While we fully acknowledge the process of understanding
based on analogies as an important method of scientific progress, we
emphasize that special care has to be taken in cases where the
theoretical support is sparse \cite{gallegatti.etal}. Motivated by this, the aim of the present paper is to carry out a careful
analysis of the high resolution data of the New York Stock Exchange  with
special emphasis on the effects caused by the size of the companies.

The paper is organized as follows. After the introduction of notations
in Section \ref{sec:intro}, Section \ref{sec:cap} presents the results
on the capitalization dependence of various measures of trading
activity.  In Section \ref{sec:value} we show that the distribution of
the traded values is not L\'evy stable as suggested previously
\cite{gopi.volume}. Consequently, the Hurst exponents of the related
time series exist, these are analyzed in Section \ref{sec:correl}.
We point out, that correlations in trading activity are strongly
non-universal with respect to company size, and that the Hurst
exponent of the traded value depends logarithmically on the mean
traded value per minute. Section \ref{sec:itt} deals with the time
intervals between trades and we give indications, that their
distribution is better described by a multiscaling ansatz than by gap scaling
proposed earlier \cite{ivanov.itt}. Finally, Section \ref{sec:conc} concludes.  

\section{Notations and data}
\label{sec:intro}
For a given time window size $\Delta t$, let the total traded value
(activity, flow) of the $i$th stock at time $t$ be 
\begin{equation}
f_i^{\Delta t}(t) = \sum_{n, t_i(n)\in [t, t+\Delta t]} V_i(n),
\label{eq:flow}
\end{equation} 
where $t_i(n)$ is the time when the $n$-th transaction of the $i$-th stock
takes place. This corresponds to the coarse-graining of the individual
events, or the so-called tick-by-tick data. Latter is denoted by
$V_i(n)$, this is the value traded in transaction $n$ and it is a
product of the price $p$ and the traded volume of stocks $\tilde V$,
\begin{equation}
V_i(n) = p_i(n) \tilde V_i(n).
\label{eq:v}
\end{equation}

Price usually changes only a little from trade to trade, while the
number of stocks traded in consecutive deals varies heavily. Thus, the
fluctuations and the statistical properties of the traded value $f(t)$
are basically governed by those of $\tilde V$. Price only serves as a
conversion factor to US dollars, that makes the comparison of stocks
possible. This way, one also automatically corrects the data for stock
splits. The statistical properties (normalized distribution,
correlations, etc.) are otherwise practically indistinguishable
between traded volume and traded value.

As the source of empirical data, we used the TAQ database
\cite{taq1993-2003} which records all transactions of the New York
Stock Exchange in the years $1993-2003$.

Finally, we note that throughout the paper we use $10$-base logarithms.

\section{Capitalization and basic measures of trading activity}
\label{sec:cap}

\begin{figure*}[!ht]
\centerline{\hbox{\includegraphics[width=163pt]{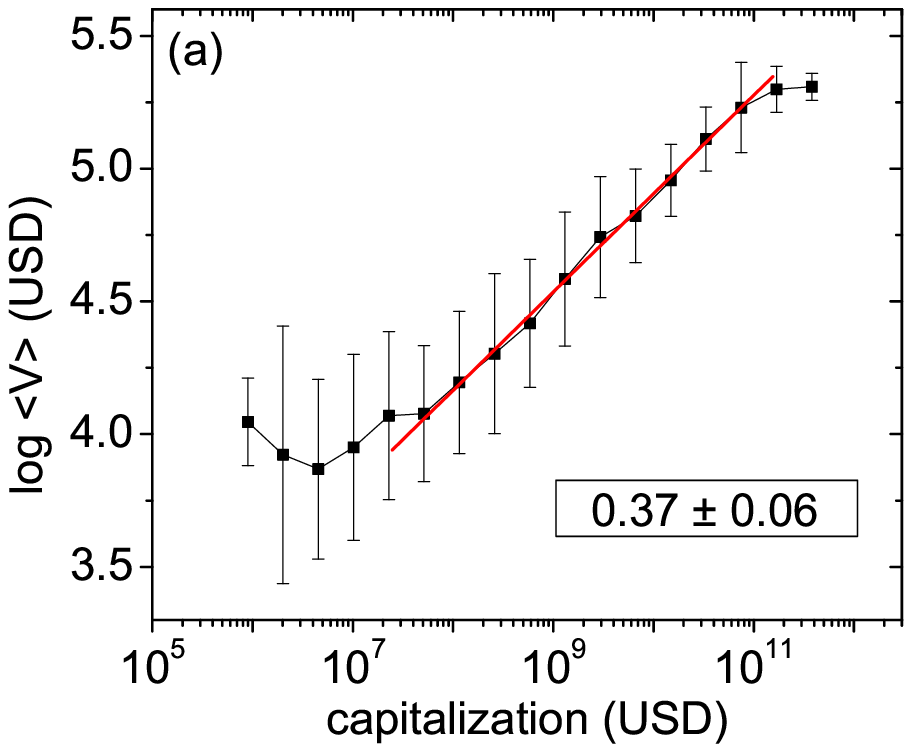}\hbox{\includegraphics[width=159pt]{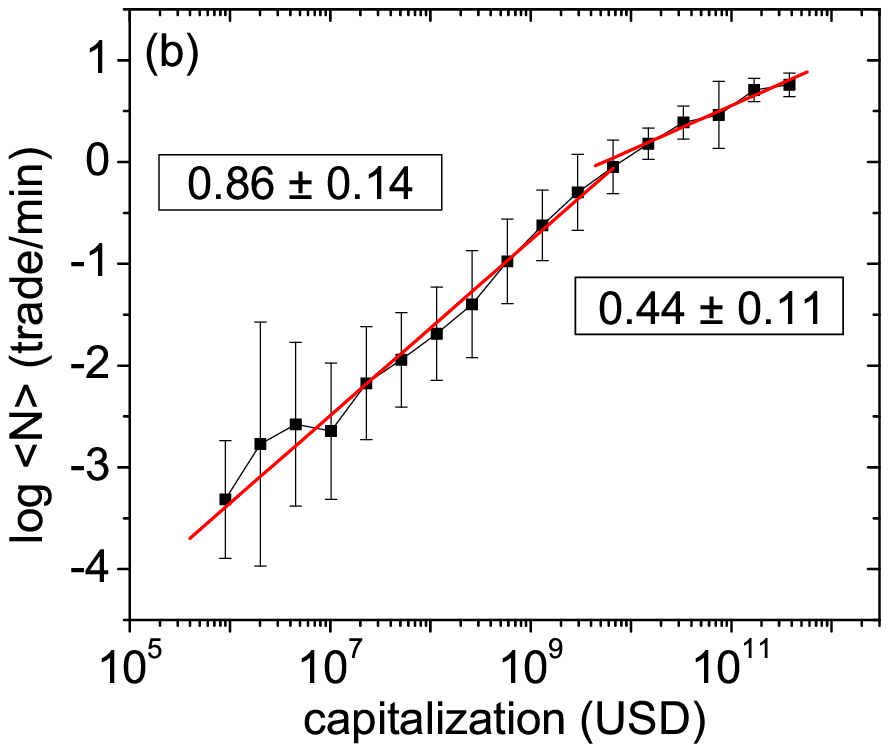}\includegraphics[width=190pt]{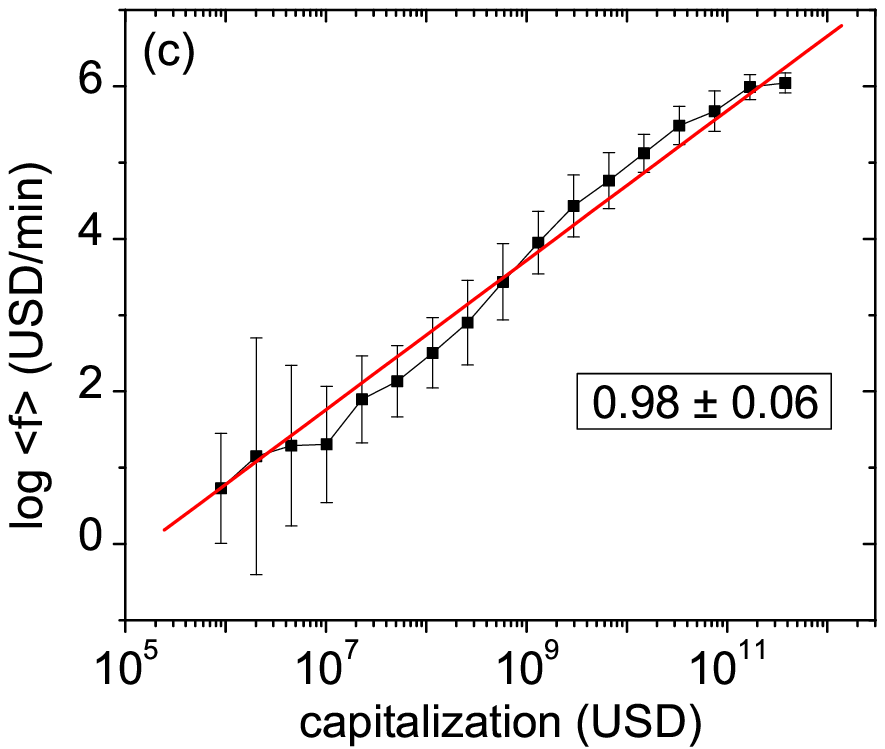}}}}
\caption{Capitalization dependence of certain measures of trading
activity in the year $2000$. The graphs are monotonically increasing and
are (piecewise) well approximated by power laws as indicated. All
three tendencies curve downward for large capitalizations. {\bf (a)}
Mean value per trade $\ev{V}$ in USD. The fitted slope corresponds to
the regime $5\cdot 10^7<C<7.5\cdot 10^{10}$ in USD. {\bf (b)} Mean
number of trades per minute $\ev{N}$. The slope on the left is from a
fit to $C<4.5\cdot 10^9$ USD, while the one on the right is for
$C>4.5\cdot 10^9$ USD. {\bf (c)} Mean trading activity (exchanged
value per minute) $\ev{f}$ in USD. The plots include
$3347$ stocks that were continuously available at NYSE during $2000$.}
\label{fig:capdep}
\end{figure*}

Many previous studies of trading focus on the stocks of large
companies. These certainly have the appealing property that price and
returns are well defined even on short time scales due to the high
frequency of trading. For infrequently traded stocks, there are extended
periods without transactions, and thus prices and returns are undefined.
In contrast, other quantities regarding the
activity of trading, such as traded value/volume or the number of
trades can be defined, even for those stocks where they are zero for
most of the time.

In this section we extend the study of Zumbach
\cite{zumbach} which concerned companies of the top two orders of
magnitude in capitalization at the London Stock Exchange. Instead, we
analyze the $3347$ stocks\footnote{Note that many minor stocks do not represent
actual companies, they are only, e.g., preferred class stocks of a larger
enterprise.} that were traded continuously at NYSE for
the year $2000$. This gives us a range of approximately $10^6\dots 6\cdot
10^{11}$ USD in capitalization.

Following Ref. \cite{zumbach}, we quantify the dependence of trading
activity on company capitalization $C_i$. Mean value per trade $\ev{V_i}$,
mean number of trades per minute $\ev{N_i}$ and mean activity (traded
value per minute) $\ev{f_i}$ are plotted versus capitalization in
Fig. \ref{fig:capdep}. Ref. \cite{zumbach} found that all three
quantities have power law dependence on $C_i$, however, this simple ansatz does
not seem to work for
our extended range of stocks. While mean trading activity can be
approximated as $\ev{f_i} \propto C_i^{0.98\pm0.06}$ to an acceptable
quality, neither $\ev{V}$ nor $\ev{N}$ can be fitted by a single power law
in the whole range of capitalization.
Nevertheless, there is -- not surprisingly -- a monotonic dependence:
higher capitalized stocks are traded more intensively.

\begin{figure}[tb]
\centerline{\includegraphics[width=185pt]{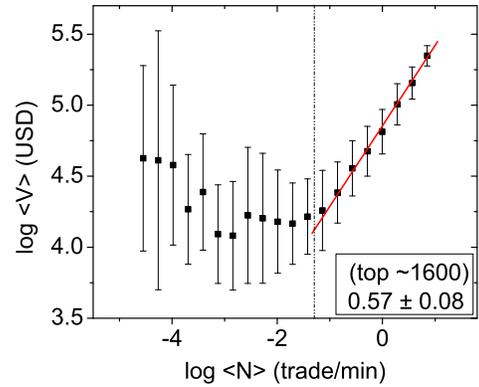}}
\caption{Plot of mean value per trade $\ev{V}$ versus mean number of
trades per minute $\ev{N}$ for the year $2000$ of NYSE. For smaller
stocks there is no clear tendency. For the top $\sim1600$ companies
($\ev{N} > 0.05$ trades/min), however, there is scaling with an
exponent $\beta = 0.57 \pm 0.08$. \emph{Note}: The plot includes
$3347$ stocks that were continuously available at NYSE during $2000$.
{\it Note:} The first few points correspond to stocks that are traded
less than daily. These typically do not represent individual companies
and might be traded according to different rules. However, unlike prices or returns,
$V$, $N$ and $f$ still remain well-defined quantities for such stocks.}
\label{fig:mNvsmV}
\end{figure}

One can gain further insight from Fig. \ref{fig:mNvsmV}, which shows,
that for the largest $1600$ stocks
\begin{equation}	
\ev{V_i} \propto \ev{N_i}^\beta
\label{eq:vvsn}
\end{equation}
with $\beta = 0.57 \pm 0.09$. The estimate based on the results of Zumbach \cite{zumbach} for the stocks in London's FTSE-100, is $\beta \approx 1$. Similar results were recently obtained for NASDAQ \cite{eisler.unified}.

For the smaller stocks there is no clear tendency. This effect can be
interpreted as follows. As we move to stocks with smaller and smaller
capitalization, the average transaction size $\ev{V}$ cannot decrease indefinitely.
Transaction costs must impose a minimal number/value of stocks in a single transaction
that can still be exchanged profitably. This minimal size is observed as the constant regime
for small $\ev{N}$. On the other hand, once a stock is
exchanged more frequently (the crossover happens at about $\ev{N} =
0.05$ trades/min), it is no more traded in this "minimal" unit. With
the growing speed of trading, trades tend to "stick together", it is
possible to exchange larger packages. This increase is clear, but not
dramatic, it is up to one order of magnitude. Although increasing
package sizes reduce transaction costs, price impact
\cite{gabaix.powerlaw, plerou.powerlaw, farmer.powerlaw,
farmer.whatreally} increases, possibly decreasing profits and thus
limiting package sizes. The interplay of these two effects has
a role in the formation of relationship \eqref{eq:vvsn}.

\section{Traded value distributions revisited}
\label{sec:value}
The statistical properties of the trading volume of stocks has
previously been investigated in Ref. \cite{gopi.volume}. That work
finds that the cumulative distribution of traded volume in $\Delta t = 15$ minute
windows has a power-law tail with a tail exponent $\lambda = 1.7 \pm
0.1$. This is the so called \emph{inverse half cube law}. Formally,
this corresponds to

\begin{equation}
{\mathbb P}_{\Delta t}(f) \propto f^{-(\lambda + 1)},
\label{eq:pl}
\end{equation}
where $\mathbb P_{\Delta t}$ is the probability density function of
traded volume (value) on a time scale $\Delta t$.

Ever since, great effort was devoted to explain this exponent in terms
of the \emph{inverse cube law} of stock returns
\cite{gopi.inversecube,gabaix.powerlaw,plerou.powerlaw}. However, the
exact distribution and the possible exponents are still much debated
\cite{farmer.powerlaw,queiros.volume}, and it has been shown that the shape of
such a distribution depends systematically on the capitalization of the company
\cite{lillo.variety}.

\begin{figure}[tb]
\centerline{\includegraphics[width=205pt]{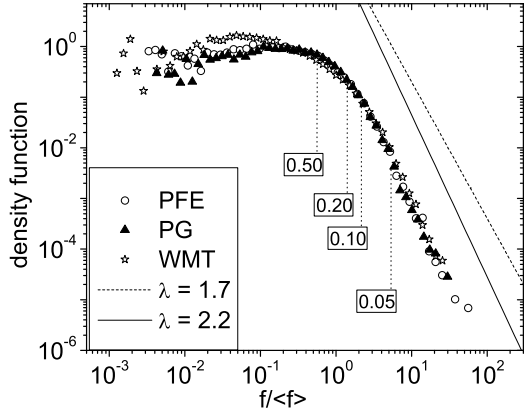}}
\caption{Distributions of traded value in $\Delta t = 15$ min time
windows, divided by the mean. The plot displays three example stocks
for the period $1994-1995$. The numbers show some upper quantiles of
the distribution (probability of values higher than indicated by the
corresponding dashed line). The dashed and solid diagonal lines represent
power-laws with exponents corresponding to $\lambda = 1.7$ and $2.2$, respectively.}
\label{fig:examplehist}
\end{figure}

The estimation of the tail exponent is a delicate matter. 
Following the methodology of Ref. \cite{gopi.volume} -- and for the
same $1994-1995$ period of data -- we repeated these measurements. Our
results for the $\Delta t = 15$ min distribution are shown in
Fig. \ref{fig:examplehist} for three majors stocks. The tails of these
distributions can be fitted by a power law over an order of magnitude,
for the top $5-10\%$ of the events. The exponent $\lambda$ we find,
is significantly higher than $1.7$, it is around $2.2$ for these
examples.

For systematic calculations of $\lambda$, there is a range of
mathematical tools available. We used three variants of Hill's method
\cite{hill,alves} to estimate the tail exponent, details can be found
in Appendices \ref{app:hill} and \ref{app:alves}. All three have a
common parameter: the number $k$ of largest events that belong to the
tail. The statistical weight associated with the tail events is
$p=k/L$, where $L$ is the total length of our time series. From
Fig. \ref{fig:examplehist} one can see, that $p\approx 5-10\%$ is the
proper choice as a threshold for the asymptotic regime.

For the two-year period $1994-1995$ and separately for the single year
$2000$, we took the $1000$ stocks with the highest total traded value
in the TAQ database. We detrended their trading activity by the well
known $U$-shaped intraday pattern (see, e.g., Ref. \cite{eisler.non-universality}). Then, we calculated the
distribution of $\lambda$ over these stocks. The median and the width
of this distribution (characterized by the half distance of the $25\%$
and $75\%$ quantiles) is shown in Tables \ref{tab:DETRlambda94-95} and
\ref{tab:DETRlambda00} for various time windows $\Delta t$.

The choice $p=0.06$ in Hill's method provides results in line with
Ref. \cite{gopi.volume}. For $\Delta t = 15$ min time windows, one
finds $\lambda = 1.71 \pm 0.20$ for the period $1994-1995$. However,
other estimates are significantly higher, $\lambda > 2$. Moreover, two
estimators show a strong tendency of increasing $\lambda$ with
increasing time windows. Monte Carlo simulations on surrogate datasets
show, that this is beyond what could be explained by decreasing sample
size. It is well known, that for $\lambda < 2$ the distribution would
have to converge to the corresponding Levy distribution when $\Delta t
\rightarrow \infty$. The measured $\lambda$'s should also be
independent of $\Delta t$. On the other hand, for $\lambda > 2$, the
$\Delta t\rightarrow \infty$ limit distribution is a
Gaussian. Accordingly, for finite samples, the measured effective
value of $\lambda$ increases with $\Delta t$. This systematic dependence
makes us conclude that there is a strong indication for the existence of the second moment.

One must keep in mind, that all three methods \emph{assume} that the
variable is asymptotically distributed as \eqref{eq:pl} and none of
them \emph{proves} it. If this does not hold, then the estimates of
exponents are only a parametric characterization of the unknown functional form, 
nevertheless, they do suggest that the second moments exist.
If the distribution is indeed of the limiting form \eqref{eq:pl}, then
although for short time windows ($\Delta t < 60$ min) there is a
fraction of stocks whose estimate gives $\lambda < 2$, even those display
$\lambda > 2$ for larger $\Delta t$. 

Based on these results we conclude that the second moments of the
distribution must exist for any $\Delta t$, therefore the
calculation of the Hurst exponent for the related time series is
meaningful. Similar qualitative features were found for the years $2001$ and $2002$ \cite{uponrequest}.

\begin{table*}[tbp]
	\centering
		\begin{tabular}{c||c||c|c||c|}
		$\Delta t$ & Hill's method ($p=0.06$) & Shifted Hill's $\lambda$ & Shifted Hill's $\varphi$ & Fraga Alves  ($p=0.1$)\\
		\hline\hline
		$1$ min & $1.43 \pm 0.09$ & $2.15\pm0.15$ & $3.0$ & $1.98 \pm 0.25$ \\
		$5$ min & $1.56 \pm 0.13$ & $2.29 \pm 0.25$ & $2.8$ & $2.04 \pm 0.25$ \\	
		$15$ min & $1.71 \pm 0.20$ & $2.55 \pm 0.35$ & $2.8$ & $2.1 \pm 0.3$ \\	
		$60$ min & $2.06 \pm 0.30$ & $2.85 \pm 0.45$ & $1.8$ & $2.1 \pm 0.4$ \\			
		$120$ min & $2.3 \pm 0.4$ & $3.15 \pm 0.70$ & $1.6$ & $2.1 \pm 0.4$ \\	
		$390$ min & $2.7 \pm 0.6$ & $3.7 \pm 0.9$ & $1.2$ & no estimate \\	
		\hline
		\end{tabular}
	\caption{Median of the tail exponents of traded value calculated by three methods for $1994-1995$. The width of the distributions is given with the half distance of the $25\%$ and $75\%$ quantiles.}
	\label{tab:DETRlambda94-95}
\end{table*}

\begin{table*}[tbp]
	\centering
		\begin{tabular}{c||c||c|c||c|}
		$\Delta t$ & Hill's method ($p=0.06$) & Shifted Hill's $\lambda$ & Shifted Hill's $\varphi$ & Fraga Alves ($p=0.1$)\\
		\hline\hline
		$1$ min & $1.63 \pm 0.13$ & $2.40 \pm 0.23$ & $2.6$ & $2.16 \pm 0.25$ \\
		$5$ min & $1.91 \pm 0.25$ & $2.8 \pm 0.5$ & $2.4$ & $2.30 \pm 0.35$ \\	
		$15$ min & $2.15 \pm 0.40$ & $3.1 \pm 0.6$ & $2.0$ & $2.35 \pm 0.40$ \\	
		$60$ min & $2.6 \pm 0.5$ & $3.45 \pm 0.8$ & $1.2$ & $2.2 \pm 0.4$ \\			
		$120$ min & $2.8 \pm 0.6$ & $3.8 \pm 1.1$ & $1.2$ & no estimate \\	
		$390$ min & $3.2 \pm 1.0$ & $5.1 \pm 0.8$ & $1.6$ & no estimate \\	
		\hline
		\end{tabular}
	\caption{Median of the tail exponents traded value calculated by three methods for $2000$. The width of the distributions is given with the half distance of the $25\%$ and $75\%$ quantiles.}
	\label{tab:DETRlambda00}
\end{table*}

\section{Non-universality of correlations in traded value time series}
\label{sec:correl}
Scaling methods \cite{vicsek.book, dfa.intro, dfa} have long been used
to characterize a wide variety of time series, including stock prices
and trading volumes \cite{bouchaud.book, stanley.book}. In particular,
the Hurst exponent $H(i)$ is usually calculated. For the traded value time series $f_i^{\Delta t}(t)$ of stock $i$, it is defined as
\begin{equation}
\label{eq:hurst}
\sigma_i^2(\Delta t) = \ev{\left (f_i^{\Delta t}(t)-\ev{f_i^{\Delta t}(t)} \right )^2}\propto\Delta t^{2H(i)},
\end{equation}
where the average is taken over the time variable $t$.
As discussed in Sec. \ref{sec:value}, the variance on the left hand
side exists for any stock or time scale $\Delta t$.

Ref. \cite{gopi.volume} finds strong correlations in $\sqrt{f_i^{\Delta t}(t)}$ with $H \approx 0.83$. Their analysis comprises the $1000$ largest companies in the period $1994-1995$ and they use $\Delta t > 1$ day except for some very
frequently traded stocks.

\begin{figure}[htb]
\includegraphics[height=180pt]{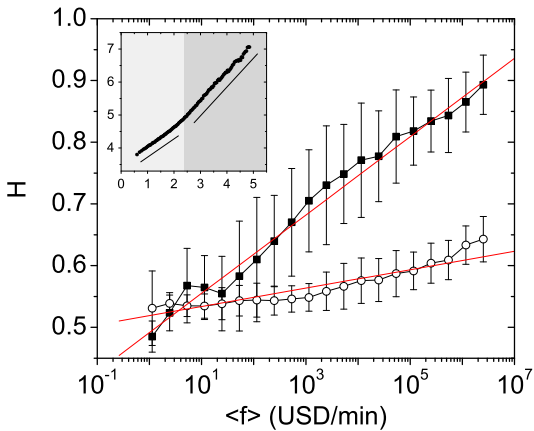}
\caption{The Hurst exponent of traded value $f$ shows logarithmic
dependence on the average traded value per minute $\ev{f}$. For
intraday fluctuations ($\Circle$), correlations in $\ev{f}$ are weak,
$H\approx 0.5 - 0.6$, the fitted slope is $\gamma(\Delta t <
\mathrm{250\space min})=0.016\pm 0.001$. Beyond the daily scale
($\blacksquare$) the effect increases: the smallest stocks show almost
no correlation ($H\approx 0.5$), while large ones display strong
persistence ($H\approx 0.9$). The fitted slope is $\gamma (\Delta t >
\mathrm{630\space min})=0.063\pm 0.002$. The inset shows the two
regimes of correlation strength for the single stock Wal-Mart (WMT) on
a log-log plot of $\sigma(\Delta t)$ versus $\Delta t$. The slopes corresponding to Hurst exponents are
$0.65$ and $0.8$.}
\label{fig:hurst}
\end{figure}

We extend these measurements to all $2647$ stocks that were continuously
traded in the period $2000-2002$. The time series display a crossover from a lower to a higher value of $H(i)$
around the time scale of one day (for an example, see the inset of
Fig. \ref{fig:hurst}). A similar effect was reported for intertrade
times of large companies \cite{ivanov.itt}. Intraday correlations are
not meaningful for some of the smallest companies as their shares are
often not exchanged for several days. Nevertheless, for any choice of
time windows, one recovers a tendency: With the change
of average traded value $\ev{f_i}$, there is a clear logarithmic trend
in the Hurst exponent, especially above the daily scale:
\begin{eqnarray}
H(i) = H(i=1) + \gamma \log \ev{f_i},
\label{eq:hurst_scaling}
\end{eqnarray}
where normalization is so that $\ev{f_{i=1}} = 1$. Measurement results and values of $\gamma$ are given in Fig. \ref{fig:hurst}. Calculations for the periods $1994-1995$ and $1998-1999$ show qualitatively
similar properties. On the grounds of a new type of scaling law \cite{eisler.non-universality},
this effect can be predicted analytically \cite{eisler.unified}. Here
we only focus on the description of the phenomenon.

Trading activity of very small stocks shows nearly no persistence. Even for $\Delta t > 1$ day,
$H\approx 0.5$. This changes as one moves to larger and larger companies. Their trading can be more correlated in the regime $\Delta t > 1$ day, up to $H\approx 0.9$. This is a clear sign of non-universality. The very nature of trading differs for different company sizes and statistics such as "distributions of Hurst
exponents" is meaningless. No typical value exists, the trend is
systematic and continuous. As Hurst exponents are closely related to
the multifractal spectra \cite{vicsek.book,ball.falpha} of $f$, those cannot be universal
either. This raises doubts about an "average multifractal spectrum" as
calculated in, e.g., Ref. \cite{drozdz.average}.

Systematic dependence of the exponent of the power spectrum of the
number of trades on capitalization was previously reported in
Ref. \cite{bonanno.dynsec}, based on the study of $88$ stocks.  This
quantity is closely related to the Hurst exponent for the time series
of the number of trades per unit time (see Ref. \cite{ivanov.itt}).
Direct analysis finds a strong dependence of
the Hurst exponent of $N$ on $\ev{N}$, but no such clear logarithmic
trend as Eq. \eqref{eq:hurst_scaling} \cite{uponrequest}.

\section{Multiscaling distribution of intertrade times}
\label{sec:itt}

Finally, we analyzed the intertrade interval series
$T_i(n=1\dots N_i-1)$, defined as the time spacings between the
$n$'th and $n+1$'th trade \cite{scalas.anomalous}. $N_i$ is the total number of trades for
stock $i$ during the period under study.

Previously, Ref. \cite{ivanov.itt} used $30$ stocks from the TAQ
database for the period $1993-1996$ and proposed that the distribution of
$T_i$ scales with the mean $\ev{T_i}$ as 
\begin{equation}
\mathbb{P}(T ,\ev{T}) = \frac{1}{\ev{T}} F(T/\ev{T}),
\label{eq:itt_scaling}
\end{equation}
and the universal scaling function $F$ is well
modeled by a Weibull distribution of the form
\begin{equation}
F(x) = \frac{\delta}{X} \left
(\frac{x}{X}\right )^{\delta - 1}\exp \left [ -\left ( \frac{x}{X}
\right )^\delta\right ],
\label{eq:ivanov.universal}
\end{equation}
where $X\approx 0.94$ and $\delta \approx 0.72$ for all the $30$
stocks, with some statistical deviations. 

We analyzed the data by including a large number of stocks with very
different capitalizations. First it has to be noted that the mean intertrade interval has
decreased drastically over the years. In this sense the stock market
cannot be considered stationary for periods much longer than one
year. We analyze the two year period $1994-1995$ (part of that used in
Ref. \cite{ivanov.itt}) and separately the single year $2000$. We use
all stocks in the TAQ database with $\ev{T} < 10^5$ sec, a total of
$3924$ and $4044$ stocks, respectively.

\begin{figure}
\centerline{\includegraphics[height=195pt]{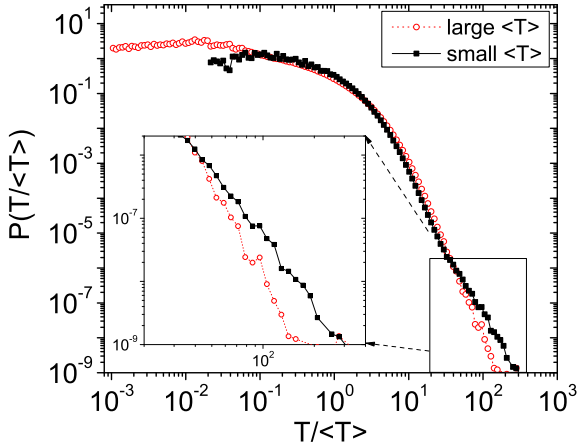}}
\caption{The distribution of $T/\ev{T}$ in the year $2000$ for two
groups of stocks with different mean intertrade times $\ev{T}$. The
group with the most frequently traded stocks (blue chips) has a
considerably greater weight for waiting times. This implies that the
distribution $\mathbb P(T, \ev{T})$ may not be universal.}
\label{fig:itt_commonhist}
\end{figure}

In order to check the validity of the gap scaling formula,
we divided the stocks into two groups\footnote{The groups were
constructed to have an approximately equal total number of
trades. Small $\ev{T}$ (top $246$ stocks): $6.48$ sec $< \ev{T} < 47.8$ sec (other $3797$ stocks), large $\ev{T}$: $47.8$ sec $< \ev{T} < 10^5$ sec.} with respect to $\ev{T}$. Then, we generated the distribution of
$T/\ev{T}$ for the groups, a comparison for the year $2000$ is shown
in Figure \ref{fig:itt_commonhist}. This already raises doubts about
the generality of Eq. \eqref{eq:ivanov.universal}: The tails of the
distribution seem to possess more weight for the group with small
$\ev{T}$ (blue chips). The direct visual comparison
of these distributions is, however, not always a reliable method to
evaluate universality. Instead, we take a less arbitrary, indirect
approach.

The consequence of the universal distribution \eqref{eq:itt_scaling}
would be that the moments of $T$ should show gap scaling: The difference
between the exponents of the $q$-th and $q+1$-th moments is
independent of $q$.
\cite{halsey.strange,murthy.sinai}:
\begin{eqnarray}
\ev{T_i^q} = C(q) \ev{T_i}^{-\tau(q)}, \label{eq:itt_multiscaling}
\end{eqnarray}
with a scaling function\footnote{We keep the negative sign to conform
with usual conventions.} $-\tau(q) \equiv q$.

\begin{figure}
\centerline{\includegraphics[height=195pt]{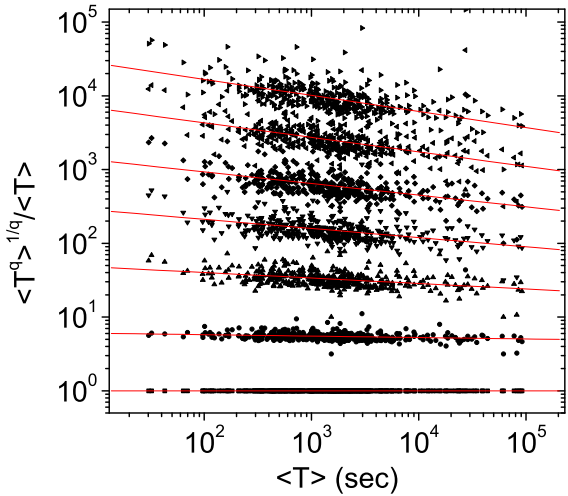}}
\caption{Scaling of integer moments of $T$, $q=1, 2, 4, 6, 8, 12, 16$ (increasing from bottom to top). The plot shows $\ev{T^q}^{1/q}/\ev{T}$, the slopes correspond to $-\tau(q)/q-1$. If the normalized distribution of $T$ were universal, the points would align on horizontal lines. \emph{Note}: The points were shifted vertically for better visibility. Only $400$ points are shown per moment, the sample period was $1994-1995$.}
\label{fig:itt_momentscaling1}
\end{figure}

\begin{figure}
\centerline{\includegraphics[height=165pt]{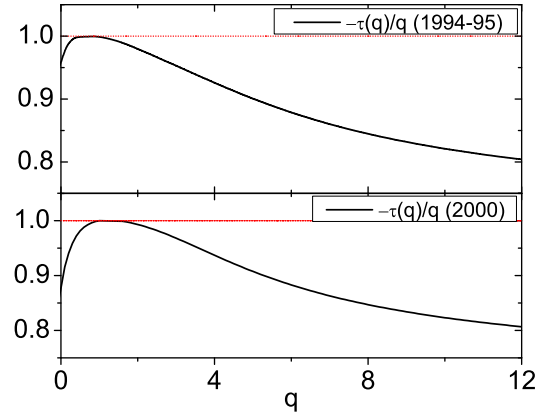}}
\caption{Scaling exponents for the moments of intertrade interval
distributions defined in Eq. \eqref{eq:itt_multiscaling}. The values
$-\tau(q)\equiv q$ would imply a universal distribution that is
independent of stock. The fact that $-\tau(q)/q<1$, shows less
frequently traded stocks display relatively lower variations in their
trading dynamics. For large $q$, the effect increases monotonically
with $q$. This suggests a difference between small and large stocks in
the tail of the distribution, which corresponds to longer periods of
inactivity.}
\label{fig:itt_momentscaling2}
\end{figure}

Instead, we find a systematic dependence of $-\tau$ on $q$, see
Fig. \ref{fig:itt_momentscaling1} for several examples of fitting and
Fig. \ref{fig:itt_momentscaling2} for all results. 
There is good fit to a power law of
type \eqref{eq:itt_multiscaling} for $4$ orders of magnitude in
$\ev{T}$ with non-trivial exponents.

The intuitive meaning of $-\tau(q \gg 1) < q$ is simple: Intertrade
times of larger (more frequently traded) stocks exhibit larger
relative fluctuations. In line with our observation from Figure
\ref{fig:itt_commonhist}, this difference must come from the tail of
the distribution, as the deviation becomes more pronounced for higher
moments.

The absence of simple universal scaling raises the question of the
capitalization dependence of the Hurst exponent for the time series
$T_i$, defined analogously to Eq. \eqref{eq:hurst} as
\begin{equation}
\label{eq:hurstittdef}
\sigma_i^2(N) = \ev{\left (\sum_{n=1}^N T_i(n)-\ev{\sum_{n=1}^N T_i(n)} \right )^2}\propto N^{2H_T(i)}.
\end{equation}

\begin{figure}[tbp]
\centerline{\includegraphics[height=175pt]{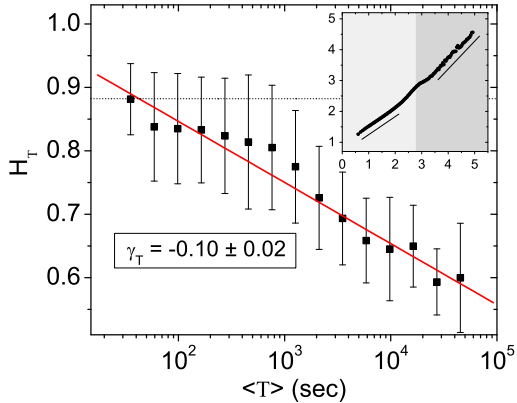}}
\caption{Hurst exponents of $T_i$ for time windows greater
than $1$ day, plotted versus the mean intertrade time
$\ev{T_i}$. Stocks that are traded less frequently, show markedly
weaker persistence of $T$ for time scales longer than $1$ day. The
dotted horizontal line serves as a reference. We used stocks with
$\ev{T} < 10^5$ sec, the sample period was $1994-1995$. The inset shows the two
regimes of correlation strength for the single stock General Electric (GE) on
a log-log plot of $\sigma(N)$ versus $N$. The slopes corresponding to Hurst exponents are
$0.6$ and $0.89$.}
\label{fig:ITT}
\end{figure}

The data show a crossover, similar to that for the traded value $f$, from a lower to a higher value of $H_T(i)$ when the window size is approximately the daily mean number of trades (for an example, see the inset of Fig. \ref{fig:ITT}).
For the restricted set studied in Ref. \cite{ivanov.itt}, the value $H_T\approx 0.94\pm0.05$
was suggested for window sizes above the crossover.

Much similarly to the case of
traded value Hurst exponents analyzed in Section \ref{sec:correl}, the
inclusion of more stocks\footnote{For a reliable calculation of Hurst
exponents, we had to discard those stocks that had less than $\ev{N} <
10^{-3}$ trades/min for $1994-1995$ and $\ev{N} < 2\cdot 10^{-3}$
trades/min for $2000$. This filtering leaves $3519$ and $3775$ stocks,
respectively.} reveals the underlying systematic
non-universality. Again, less frequently traded stocks appear to have
weaker autocorrelations as $H_T$ decreases monotonically with growing
$\ev{T}$. One can fit an approximate logarithmic law \footnote{As
intertrade intervals are closely related to the number of trades per
minute $N(t)$, it is not surprising to find the similar tendency for
that quantity \cite{bonanno.dynsec}.}$\null^,$\footnote{Note that for
window sizes smaller than the daily mean number of trades, intertrade
times are only weakly correlated and the Hurst exponent is nearly
independent of $\ev{T}$. This is analogous to what was seen for traded
value records in Sec. \ref{sec:correl}.} to characterize the trend:
\begin{equation}
H_T = H_T(\ev{T}=1)+\gamma_T\log\ev{T},
\label{eq:hurstitt}
\end{equation}
where $\gamma_T = -0.10\pm 0.02$ for the period $1994-1995$ (see
Fig. \ref{fig:ITT}) and $\gamma_T = -0.08 \pm 0.02$ for the year
$2000$ \cite{uponrequest}.

In their recent preprint, Yuen and Ivanov \cite{ivanov.unpublished} independently show a tendency similar to Eq. \eqref{eq:hurstitt} for intertrade times of NYSE and NASDAQ in a different set of stocks.

\section{Conclusions}
\label{sec:conc}

In this paper we revisited some ``stylized facts'' of stock market
data and found in several ways alterations from earlier
conclusions. The main difference in our approach was -- besides the
comparative application of extrapolation techniques -- the extension of
the range of capitalization of the studied firms. This enabled us to
investigate the dependence of the trading characteristics on
capitalization itself. In fact, in many cases we found fundamental
dependence on this parameter.

We have shown that trading activity $\ev{f}$, the number of
trades per minute $\ev{N}$ and the mean size of transactions
$\ev{V}$ display non-trivial, but monotonic dependence on company
capitalization.

We have given evidence that the distribution of traded value in fixed
time windows is not Levy stable. If a power law is fitted to the tail
of the distribution, a careful analysis yields to an exponent
$\lambda$, which is -- even for short time windows -- in most cases greater
than $2$, and then increases with increasing time window 
indicating the existence of the second moment of the distribution.
Consequently, the Hurst
exponent $H$ for its variance can be defined and it depends on the
mean trading activity $\ev{f}$ as
$$H(i) = H(i=1) + \gamma \log \ev{f_i}.$$

The mean transaction size can be fitted to a power-law dependence on
the trading frequency for moderate to large companies.

The distribution of the waiting times between trades is better
described multiscaling than by gap scaling. It is characterized by an
increase in both correlations and relative fluctuations with growing
trading frequency (i.e. increasing capitalization).

Our findings indicate that special care must be taken when concepts like
scaling and universality are applied to financial processes. The
modeling of the market should be extended to the capitalization
dependence of the characteristic quantities and this seems a
real challenge at present.

\bibliographystyle{unsrt}
\bibliography{sizematters3}

\begin{acknowledgement}
The authors are indebted to Vasiliki Plerou and Parameswaran
Gopikrishnan for discussions on Hill's method and G\'eza Gy\"orgyi for
his insights on correlated time series.  They also thank Gy\"orgy
Andor and \'Ad\'am Zawadowski for their support with the data. Finally, they would like to thank
Plamen Ch. Ivanov for calling their attention to some recent results on market dynamics.
JK is member of the Center for
Applied Mathematics and Computational Physics, BME; furthermore, he is
grateful for the hospitality of Dietrich Wolf (Duisburg) and of the
Humboldt Foundation. Support by OTKA T049238 is acknowledged.
\end{acknowledgement}

\appendix

\section{The estimation of tail exponents $\lambda$}
\label{app:hill}

In the following, for every measurement we give the median estimates
of $\lambda$ for the $1000$ stocks with highest traded value during the
investigated period. The error bars show the half distance between the
$25\%$ and $75\%$ quantiles of $\lambda$.

\subsection{Hill's estimator}

Hill's estimator \cite{hill} is a statistically consistent method to
estimate the tail exponent $\lambda$ from random samples taken from a
distribution that asymptotically has the power-law form
\eqref{eq:pl}. The procedure first sorts the sample $f(t=1\dots L)$ in
decreasing order. We are going to denote this series by $f[t]$, so
that $f[1]>f[2]>f[3]>\dots$. Then, one defines the tail of the
distribution by setting an arbitrary number $k$ of points to be
included in the estimation procedure. The estimate of the inverse tail
exponent is
\begin{equation}
\hat \lambda^{-1}(k) = \left [ \frac{1}{(k-1)}\sum_{t=1}^{k-1}\log f[t]\right ] - \log f[k],
\label{eq:hill}
\end{equation}
given that $k\rightarrow \infty$ and
$p=k/L\rightarrow 0$. If the sampled distribution is of the form
\eqref{eq:pl}, then by increasing $k$, the estimator converges rapidly
to the actual value of $\lambda^{-1}$. However, in the case of traded
value data, this turns out not to be the case.

The inset of Fig. \ref{fig:hill}(a) -- a so called Hill plot -- shows,
that there is a systematic dependence of $\lambda$ on $p$ and no
convergence is observed. With the inclusion of less tail events, the
exponent increases sharply, beyond the $\lambda = 2$ threshold for
L\'evy stability. Further evidence for the lack of L\'evy stability is
that on increasing the time scale $\Delta t$, the estimated tail
exponents also increase further as shown in Fig. \ref{fig:hill}(a).

This type of behavior is not new to mathematical statistics (see,
e.g., Ref. \cite{alves}). It is possible, that the distribution decays
faster than a power law and thus no finite $\lambda$
exists. Alternatively, the power law may not be centered around zero,
but instead it can be of the form
\begin{equation}
{\mathbb P}_{\Delta t}(f) \propto \left (f+f_0\right)^{-(\lambda + 1)}.
\label{eq:shiftedpl}
\end{equation}
In this latter case, there is a finite $\lambda$, but as the sample size $T$
is usually too small, the estimator displays the above bias. One can
either try to approximate the value of $f_0$ and shift the data
accordingly, so that Hill's estimator converges properly, or try to
find another estimator that is insensitive to this shifting constant.

We have tried both approaches and they yielded qualitatively similar results.

\subsection{Shifted Hill's estimator}

\begin{figure*}[tbp]
\centerline{\includegraphics[height=165pt]{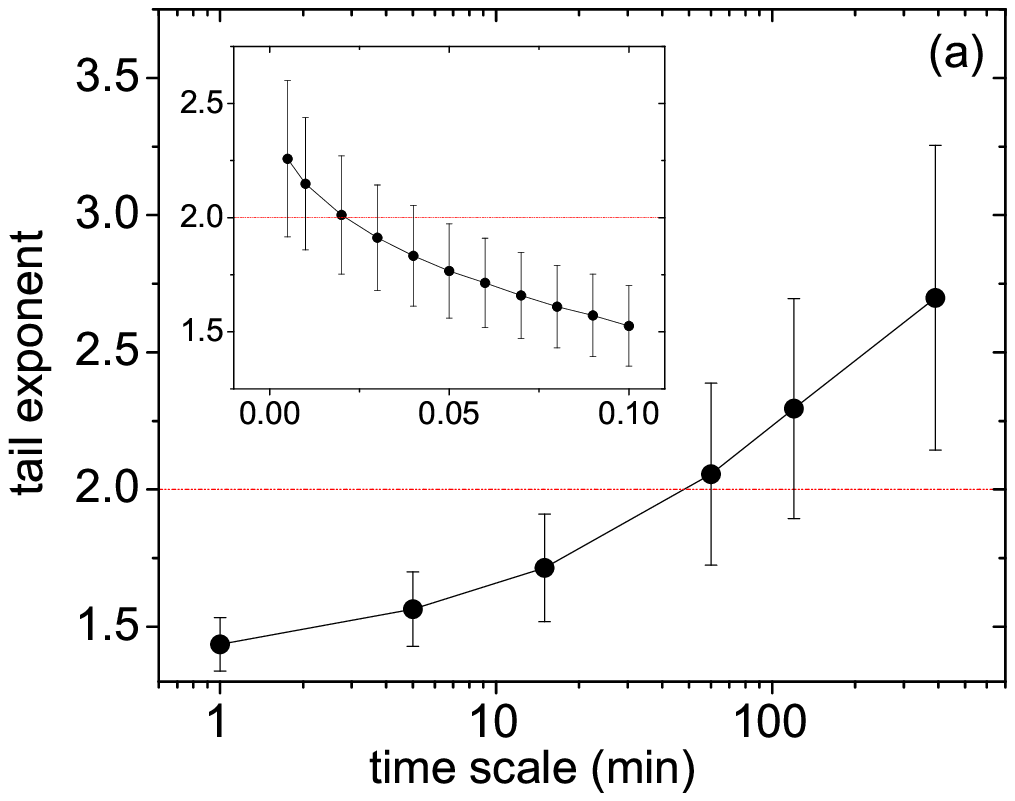}\hskip10pt\includegraphics[height=165pt]{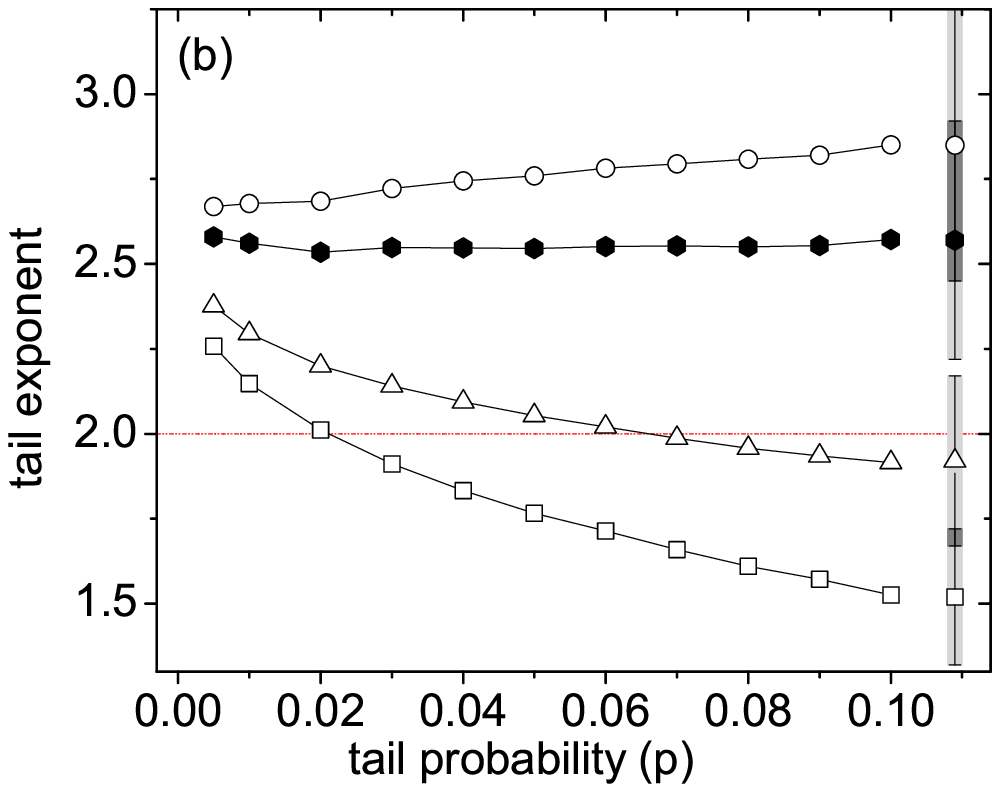}}
\centerline{\includegraphics[height=165pt]{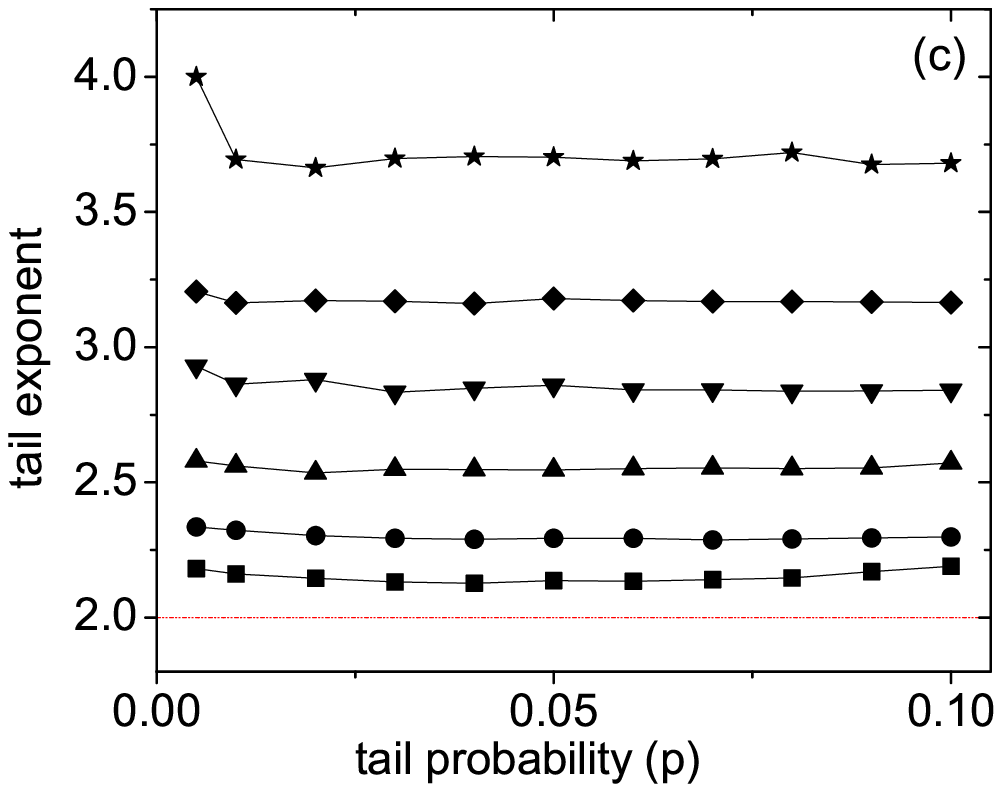}\hskip24pt\includegraphics[height=165pt]{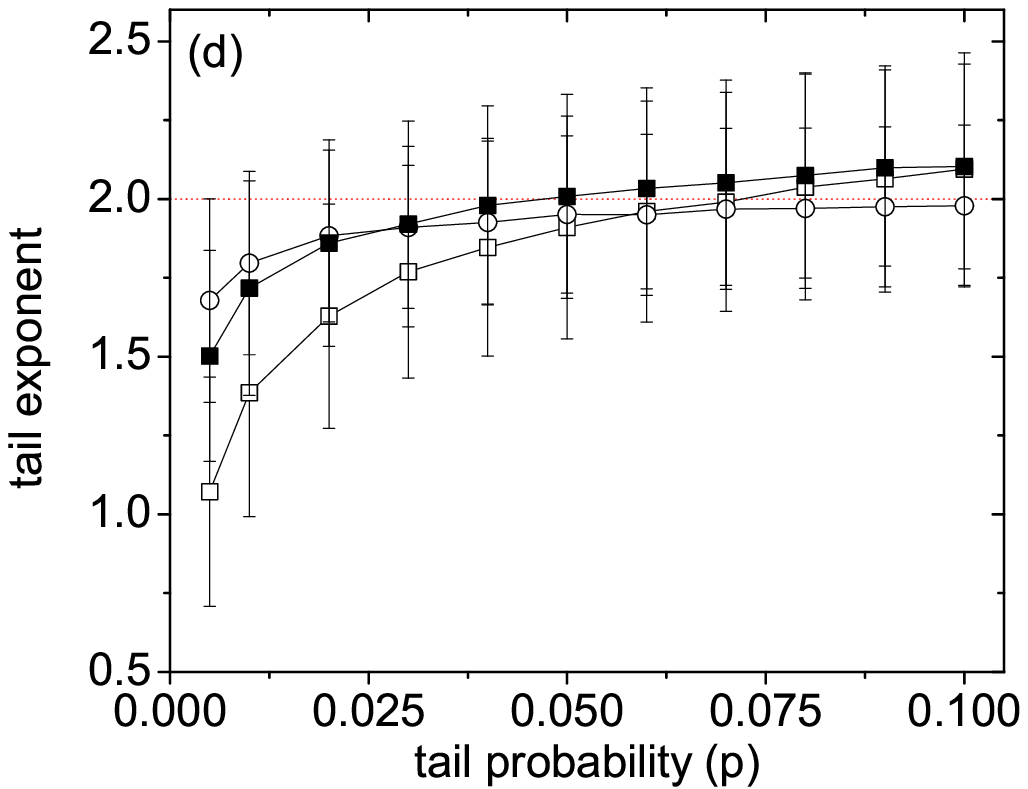}}
\caption{{\bf (a)} Hill' s estimates of $\lambda$ for different sizes
of the time window with the tail probability set as $p = 0.06$. The
monotonic trend indicates that the distribution is not be Levy
stable. The inset shows, that for $\Delta t = 15$ min the effective
tail exponent $\lambda$ depends monotonically on the choice for tail
probability $p$. Thus, Hill's estimates are unreliable, because they
depend strongly on an arbitrary parameter. {\bf (b)} Dependence of the
Hill plots for $\Delta t = 15$ min on the shifting constant
$\varphi$. The values of $\varphi$ from bottom to top: $0$ ($\Box$),
$1$ ($\bigtriangleup$), $2.8$ ($\CIRCLE$, optimal shift), $3.0$
($\Circle$). Typical error bars are given on the right, darker gray indicates the regimes where they overlap. 
{\bf (c)} Hill plots of the optimally shifted Hill's estimators
for various time windows. The values of $\Delta t$ from bottom to top:
$1$ min ($\blacksquare$), $5$ min ($\CIRCLE$), $15$ min
($\blacktriangle$), $60$ min ($\blacktriangledown$), $120$ min
($\blacklozenge$), $390$ min ($\bigstar$). One finds $\lambda > 2$ and
the strong increasing tendency in $\lambda$ with $\Delta t$ implies
that the distribution is not Levy stable. {\bf (d)} Hill plots of the
Fraga Alves estimator for three time window sizes $\Delta t$: $1$ min
($\Circle$), $15$ min ($\blacksquare$), $60$ min ($\Box$). The method
gives a lower estimate of $\lambda \approx 2$.}
\label{fig:hill}
\end{figure*}

One can apply Hill's estimator to the points $f[t=1\dots
L]+\varphi\ev{f}$, where $\varphi$ is a constant parameter and look
for a value, where the estimator $\lambda(k)$ becomes independent\footnote{More precisely, we increased $\varphi$ from $0$ by increments of $0.2$ and looked for $\lambda(k, \varphi) \approx \lambda(\varphi)$. The method is very sensitive to the
proper choice of $\varphi$. For high values of $\Delta t$, there is a
low number of data points, and the estimates of $\lambda$ may be very
noisy. In this case we chose $\varphi$, where the estimate of $\lambda$ is lower.} of
$k$, i.e., Hill's estimator truly finds a power-law decay that is now
consistent with Eq. \eqref{eq:shiftedpl}. For an example, see Fig. \ref{fig:hill}(a). This happens, when
$\varphi\ev{f} = f_0$. How this shift by $\varphi\ev{f}$ affects the
Hill plots is shown in Fig. \ref{fig:hill}(c) for the case of $\Delta
t = 15$ min. One finds, that in this case $\varphi \approx 2.8$ gives
reasonable results, while $\lambda = 2.55 \pm 0.35$. One can repeat the
procedure for various time scales $\Delta t$. The median Hill plots
are shown in Fig. \ref{fig:hill}(d), while $\lambda(\Delta t)$ and
$\varphi(\Delta t)$ are given in Table
\ref{tab:DETRlambda94-95}. Again, one finds a significant increase of
the tail exponent with growing $\Delta t$. This underlines our
previous expectation that traded value distributions are not L\'evy
stable and thus have a finite variance.

\subsection{Fraga Alves estimator}

A more sophisticated approach to estimate tail exponents of
distributions of the type \eqref{eq:shiftedpl}, is a recent variant of
Hill's method, proposed by Fraga Alves \cite{alves}. The algorithm is
described in detail in Appendix \ref{app:alves} and its estimates of
$\lambda$ are -- in an exact mathematical sense -- independent of the
shift $f_0$ present in the density function, unlike those of the original
Hill's estimator \eqref{eq:hill}.

We applied the estimator to the same dataset, the Hill plots for
$\Delta t = 1, 15, 60$ min are shown in Fig. \ref{fig:hill}(d). What
one finds is a very different behavior from the shifted Hill's
estimator. The estimate of $\lambda$ increases with growing $p$, i.e.,
the more points included. This is due to that the Fraga Alves
estimator converges much slower than Hill's estimator, and -- as
Fig. \ref{fig:hill}(d) and Monte Carlo simulations on surrogate
datasets indicate -- it converges from below. On the other hand,
setting the threshold as high as $p=0.1$ may include events that no
more belong to the power law regime, which also results in a reduced,
effective exponent due to the shape of the distribution, shown in
Fig. \ref{fig:examplehist}. Consequently, this method provides a lower
estimate of $\lambda$. Still, the calculated values are mostly above
$2$. Finally, one must note that for $\Delta t \geq 120$ min, the
number of points was inadequate to provide any proper estimate at all.

\section{The algorithm of the Fraga Alves estimator}
\label{app:alves}
Ref. \cite{alves} describes a method to approximate the parameter
$\lambda$ from a sample of a random variable that is asymptotically
distributed as $$\mathbb{P}_{\Delta t}(f)\propto (f+f_0)^{-(\lambda +
1)}.$$ First, one sorts the sample $f(t=1\dots L)$ in decreasing
order. We denote this series by $f[t]$, so that
$f[1]>f[2]>f[3]>\dots$. Then, the procedure consists of the five steps
formulated below:

\begin{enumerate}

\item $k_0^* = 2k^{2/3}$
\item $$\hat \lambda^{-1}(k_0^*, k) =
\frac{1}{k_0^*-1}\sum_{t=1}^{k_0^*-1} \log
\frac{f[t]-f[k]}{f[k_0]-f[k]}$$
\item $$k_0 = C_0^{1/(2\hat \lambda^{-1}(k_0^*, k)+1)}k^\alpha, $$
where
$$C_0 = \frac{(1+\hat \lambda^{-1}(k_0^*, k))^2}{2\hat \lambda^{-1}(k_0^*, k)},$$ and
$$\alpha = \frac{2\hat \lambda^{-1}(k_0^*, k)}{2\hat \lambda^{-1}(k_0^*, k)+1}.$$
\item $$\hat \lambda^{-1}(k_0, k) = \frac{1}{k_0-1}\sum_{t=1}^{k_0-1}
\log \frac{f[t]-f[k]}{f[k_0]-f[k]}$$
\item Finally, the estimate of the inverse tail exponent is given by
$$\lambda^{-1}(k_0, k) = \hat \lambda^{-1}(k_0, k) -
\sqrt{\frac{\hat \lambda^{-1}(k_0, k)}{2k_0}}.$$ 
$\lambda^{-1}(k_0, k)$ converges to the inverse tail exponent, if $L \rightarrow \infty$, $k/L \rightarrow 0$ and $k_0/k \rightarrow 0$.

\end{enumerate}

\end{document}